\documentclass[journal]{IEEEtran}
\ifCLASSINFOpdf
\else
\fi
\def\bb0{{\mathbb{0}}}

\def\bb{{\boldsymbol{b}}}

\def\bv{{\boldsymbol{v}}}

\def\bx{{\boldsymbol{x}}}
\def\by{{\boldsymbol{y}}}

\def\b0{{\boldsymbol{0}}}

\def\bC{{\boldsymbol{C}}}
\def\bD{{\boldsymbol{D}}}

\def\bF{{\boldsymbol{F}}}

\def\bH{{\boldsymbol{H}}}

\def\bT{{\boldsymbol{T}}}

\def\b{{\mathrm{b}}}

\def\r0{{\mathbf{0}}}

\def\cH{\mathcal{H}}

\def\cO{\mathcal{O}}

\def\bsf0{{\bm{\mathsf{0}}}}

\def\Nt{{N_{\mathrm{t}}}}
\def\Nr{{N_{\mathrm{r}}}}

\def\Nmax{{N_{\mathsf{max}}}}

\def\Pt{{P_{\mathrm{t}}}}

\def\N0{{N_{\mathrm{0}}}}

\def\bsf{{\boldsymbol{s}_\mathrm{f}}}

\newcommand{\be}{\begin{equation}}
\newcommand{\ee}{\end{equation}}
\newcommand{\bal}{\begin{align}}
\newcommand{\eal}{\end{align}}

\def\SNR    {{\mathsf{SNR}}}

\def\Gt {G_{\mathrm{t}}}
\def\Gr {G_{\mathrm{r}}}

\def\thetat{\theta_{\rm t}}
\def\thetar{\theta_{\rm r}}
\def\phir{\phi_{\rm r}}
\def\Nmin{N_{\sf min}}
\def\Nmax{N_{\sf max}}

\usepackage{graphicx}
\usepackage{amsmath}
\usepackage{subfigure}
\usepackage{mathtools}
\usepackage{epsfig}
\usepackage{float}
\usepackage{lipsum}
\usepackage{stfloats}
\usepackage{array}
\usepackage{amssymb}
\usepackage{cite}
\usepackage{url}
\usepackage{algorithm2e}
\usepackage{algorithmic}
\usepackage{multirow}
\usepackage{fancyh}
\usepackage[keeplastbox]{flushend}
\usepackage{upgreek}
\usepackage{dsfont}
\usepackage{gensymb}

\usepackage{multicol, blindtext}
\usepackage{mathtools, cuted}

\usepackage{color}

\newtheorem{lemma}{Lemma}

\newcommand{\dt}{{d_{\rm t}}}
\newcommand{\dr}{{d_{\rm r}}}

\DeclarePairedDelimiter\floor{\lfloor}{\rfloor}
\normalsize
\begin{document}
%
\title{Reconfigurable ULAs for Line-of-Sight \\ MIMO Transmission
\thanks{H. Do and N. Lee are with POSTECH, Korea 37673 (email: \{doheedong,nylee\}@postech.ac.kr). Their work is supported by the Samsung Research Funding \& Incubation Center of Samsung Electronics under Project SRFC-IT1702-04.
A. Lozano is with Univ. Pompeu
Fabra, 08018 Barcelona (e-mail: angel.lozano@upf.edu). His work is supported by the European Research Council under the H2020 Framework Programme/ERC grant agreement 694974 and by MINECO's Projects RTI2018-102112 and RTI2018-101040.
Parts of this paper were presented at the 2020 IEEE Int'l Symp. on Inform. Theory (ISIT'20) \cite{ISIT}.}
}

\author{\IEEEauthorblockN{Heedong~Do},
 {\it Student Member,~IEEE},
\and
\IEEEauthorblockN{Namyoon~Lee},
{\it Senior Member,~IEEE},
\and
\IEEEauthorblockN{Angel~Lozano},
{\it Fellow,~IEEE}
}
\maketitle

\maketitle

\begin{abstract}
This paper establishes an upper bound on the capacity of line-of-sight multiantenna channels over all possible antenna arrangements and shows that uniform linear arrays (ULAs) with an SNR-dependent rotation of transmitter or receiver can closely approach such capacity---and in fact achieve it at low and high SNR, and asymptotically in the numbers of antennas. Then, as an alternative to mechanically rotating ULAs, we propose to electronically select among multiple ULAs having a radial disposition at either transmitter or receiver, and we bound the shortfall from capacity as a function of the number of such ULAs. With only three ULAs, properly angled, $96\%$ of the capacity can be achieved. Finally, we further introduce reduced-complexity precoders and linear receivers that capitalize on the structure of the channels spawned by these configurable ULA architectures.

\end{abstract}



%
\IEEEpeerreviewmaketitle

\section{Introduction}

An unrelenting trend in the evolution of wireless systems is the move to ever higher frequencies, so as to exploit ever wider bandwidths.
The current frontier is at about 90 GHz, but researchers already have their eyes set on sub-terahertz bands where new applications await, including kiosk information transfers \cite{Song:18}, wireless backhaul \cite{Hur:11,Hur:13,Cvetkowski:16}, and wireless interconnections within datacenters \cite{Halperin:11}.
Another consolidated feature of wireless systems are multiple-input multiple-output (MIMO) techniques, which bring about major improvements in spectral and energy efficiency \cite{Foundations:18}.

At microwave frequencies, MIMO enables spatial multiplexing by virtue of multipath propagation, whereby the environment acts as a lens that delivers a high-rank channel \cite{Driessen:99}.
As we move up in frequency, into the mmWave realm and then into sub-terahertz territory, the transmission range necessarily shrinks and the propagation becomes mostly line-of-sight (LOS). The multipath lensing effect dwindles. At the same time, because the wavelength also shrinks dramatically, it becomes progressively possible to span a high-rank channel based only on the array apertures themselves \cite{jiang2005spherical,bohagen2009spherical}.
In particular, parallel uniform linear arrays (ULAs) can give rise to a channel with all-equal singular values, ideal for spatial multiplexing \cite{Torkildson:11}, provided that the antenna spacing is $d=\sqrt{\lambda D/N}$ where $\lambda$ is the wavelength, $D$ the transmission range, and $N$ the number of antennas at each end. With this so-called \emph{Rayleigh spacing} within the ULAs, directional signals can be launched and then resolved at the receiver without cross-talk.
And, at sub-terahertz frequencies, Rayleigh spacing become feasible within reasonably compact arrays: at $300$ GHz, for instance, a 16-antenna array with an LOS range of $D=5$ m would occupy $26.2$ cm, and far less if arranged as a two-dimensional uniform rectangular array (URA).

With spatial multiplexing as an objective, the antenna spacings that yield a channel with all-equal singular values have been determined, not only for ULAs as detailed above, but for a variety of array geometries \cite{Driessen:99, Gesbert:02, Haustein:03, Bohagen:05, Sarris:05, Bohagen:07, Sarris:07, Larsson:05, Song:151}. Moreover, the efficacy of the ensuing spatial multiplexing has been demonstrated experimentally, for now with up to $N=4$ antennas at mmWave frequencies \cite{Sheldon:081, Sheldon:082, Sheldon:09, Bao:15,Hofmann:16, Halsig:172, yan201811,sellin2019ericsson}.
Spatial multiplexing, however, is the desirable transmission strategy only at high SNR.
At low SNR, alternatively, maximizing the received power is of essence \cite[ch. 5]{Foundations:18}, and that demands beamforming over a channel whose maximum singular value is as large as possible, rather than having all-equal singular values \cite{Matthaiou:10,Chiurtu:00}.
This suggests that the ULA antenna spacings, and more generally the antenna arrangements, should depend on the SNR so as to strike a balance between spatial multiplexing and beamforming \cite{sun2014mimo}. As a step in this direction, \cite{Matthaiou:10} proposes switching, as a function of the SNR, among three ULAs with distinct antenna spacings.
Also recognizing that both spatial multiplexing and beamforming are relevant ingredients, other works such as \cite{Allerton2006,Torkildson:11,Song:152} propound the use of arrays-of-subarrays, which mix small and large antenna spacings and, advantageously, require a reduced number of radio-frequency chains \cite{Lin:16}.

The present paper shows how a single ULA can be reconfigured, through simple rotation, to closely approach the LOS capacity at any desired SNR. Precisely:

\begin{itemize}
\item An information-theoretic footing is established in the form of an upper bound on the LOS capacity over all possible antenna arrangements.
\item The ULA antenna spacings that are optimum as a function of the SNR are determined, and it is shown that ULAs with such spacings approach the LOS capacity within a gap that vanishes as the number of antennas increases, and also at low and high SNR.
\item An architecture is proposed in which ULAs can be configured without changing their antenna spacing, rather through mere rotation of transmitter or receiver, or else by electronically selecting among various ULAs in a radial disposition.
\item For such configurable architectures, low-complexity precoders and linear receivers are also put forth.
\end{itemize}

The paper is organized as follows. Section \ref{sec:LOS} introduces the LOS channel model and its capacity. Then, Section \ref{sec:ULA} specializes the channel model to parallel and non-parallel ULAs. The upper bound on the LOS capacity is presented in Section \ref{Iniesta}, and proved in an appendix. Section~\ref{ITA} sets the stage for the configurable architectures presented in Section \ref{Netflix}. Finally, the low-complexity precoders and receivers are laid down in Section \ref{sec:complexity} and the paper concludes in Section \ref{sec:conclusion}, with further proofs relegated to subsequent appendices.

\section{Channel Model and Capacity}
\label{sec:LOS}

Consider $\Nt$ transmit and $\Nr$ receive antennas connected by an LOS channel. The far-field complex baseband channel coefficient from the $m$th transmit to the $n$th receive antenna is
\begin{align}
    \label{eq:LoSSISO}
    h_{n,m} = \frac{ \sqrt{\Gt \Gr} \, \lambda }{4 \pi \, D_{n,m} } \, e^{-j\frac{2\pi}{\lambda}D_{n,m}} \qquad& n = 0,\ldots,\Nr-1\\&m = 0,\ldots,\Nt-1 \nonumber
\end{align}
where $D_{n,m} $ is the distance from the $m$th transmit to the $n$th receive antenna
while $\Gt$ and $\Gr$ are the respective antenna gains. 
Provided the antenna apertures are small relative to $D_{n,m}$, the magnitude $|h_{n,m}|$ is approximately constant across $m$ and $n$ while $D_{n,m} \approx D$ such that only the phase variations need to be modeled. These are captured by the normalized matrix
\begin{align}
\bH =
\begin{bmatrix}
     e^{-j\frac{2\pi}{\lambda}d_{0,0}}  & \cdots &  e^{-j\frac{2\pi}{\lambda}d_{0,N_{\rm t}-1}} \\
    \vdots & \ddots & \vdots \\
    e^{-j\frac{2\pi}{\lambda}d_{N_{\rm r}-1,0}} & \cdots & e^{-j\frac{2\pi}{\lambda}d_{N_{\rm r}-1,N_{\rm t}-1}}
    \end{bmatrix},
\label{eq:LoSMIMO}
\end{align}
which, letting $\sigma_n(\cdot)$ denote the $j$th singular value of a matrix, satisfies
\begin{align}
    \sum_{n=0}^{\Nmin-1} \sigma^2_n(\bH) 
     =N_{\rm r}N_{\rm t}.
\label{eq:LoSprop}
\end{align}
For the sake of compactness, we define $\Nmin=\min(N_{\rm r},N_{\rm t})$ and $\Nmax=\max(N_{\rm r},N_{\rm t})$.
We further define $\cH$ as the set of normalized matrices $\bH$ generated by all possible antenna placements that respect 
 the condition of apertures much smaller than $D$.

 
At the receiver,
  \begin{align}
  \SNR = \frac{ \lambda^2 \Gt \Gr P_{\rm t}}{(4\pi D)^2 B N_0} 
 \end{align}
where $\Pt$ is the transmit power, $B$ the bandwidth, and $N_0$ the noise spectral density. For a range $D$ and specific parameters (wavelength, antenna gains, power, and bandwidth), the SNR becomes determined. The information-theoretic capacity of a specific channel realization $\bH$ is then \cite{Foundations:18}
\begin{align}
& \!\! C(\bH,\SNR) = \!\! \max_{\substack{\sum_{n=0}^{N-1} \! p_n = \SNR \\ p_n\geq 0}} \, \sum_{n=0}^{N-1}\log_2 \! \Big(1+ p_n \, \sigma^2_n(\bH) \Big)     \\
& \quad =  \sum_{n=0}^{N-1} \log_2 \! \left(1+ \left[ \frac{1}{\gamma} - \frac{1}{\sigma^2_n(\bH)}  \right]^{+} \!\! \sigma^2_n(\bH) \right) 
\label{TerStegen}
\end{align}
with $\gamma$ such that $\sum_{n=0}^{N-1} p_n = \SNR$ and $[z]^+=\min(0,z)$.
Achieving $C(\bH,\SNR)$ requires a precoder aligned with the right singular vectors of $\bH$ and whose powers on those directions are $p_0,\ldots,p_{N-1}$, as well as a linear receiver aligned with the left singular vectors of $\bH$.

The problem of establishing the LOS capacity broadens to that of identifying the antenna placements that yield the LOS channel whose individual capacity is largest, i.e.,
\begin{align}
 C(\SNR)= \max_{\bH \in \mathcal{H}} \, C(\bH,\SNR) .
 \label{eq:optFormulation}
\end{align} 


\section{Array Structures}
\label{sec:ULA}

This section presents compact expressions for the LOS channels spawned by different ULA configurations.

\subsection{Parallel ULAs}

For parallel transmit and receive ULAs with respective antenna spacings $\dt$ and $\dr$, basic trigonometry leads to \cite{Larsson:05}
\begin{align}
    h_{n,m} =e^{-j\frac{2\pi}{\lambda}\sqrt{D^2+(n \dr- m \dt)^2}} ,
\end{align}    
which, under our proviso that the antenna apertures are small relative to $D$, satisfies
\begin{align}    
 h_{n,m}   & \approx e^{-j 2\pi \frac{D}{\lambda}} \, \underbrace{ e^{-j \pi \frac{n^2 }{\lambda D} d^2_{\rm r}} }_{\text{RX phase shifts}} \, e^{j2\pi \frac{n m }{\lambda D} \dr \dt} \, \underbrace{ e^{-j\pi \frac{m^2}{\lambda D} d^2_{\rm t}} }_{\text{TX phase shifts}} . \label{eq:ulaChannel}
\end{align}
The constant phase in the leading term and the phase shifts across the transmit and receive arrays do not affect the singular values, and can be easily compensated for, hence we
concentrate on the remaining term, which gives the Vandermonde Matrix
\begin{align}
\setlength\arraycolsep{2pt}
\bH_{\sf ULA} = \begin{bmatrix}
    e^{j2\pi\eta \frac{0\times0}{\Nmax}} & \cdots & e^{j2\pi\eta \frac{(N_{\rm t}-1)\times0}{\Nmax}} \\
    \vdots & \ddots & \vdots \\
    e^{j2\pi\eta \frac{0\times (N_{\rm r}-1)}{\Nmax}} & \cdots & e^{j2\pi\eta \frac{(N_{\rm t}-1)\times(N_{\rm r}-1)}{\Nmax}}
    \end{bmatrix}
    \label{Messi}
\end{align}
where we have introduced 
\be
\eta=\frac{ \dr \dt \Nmax }{\lambda D}  
\label{TopoMates}
\ee
as a parameter that compactly describes the parallel ULA configuration. 
Rayleigh antenna spacings correspond to $\eta=1$, whereby the matrix becomes column-orthogonal if $N_{\rm r}\geq N_{\rm t}$ and row-orthogonal if $N_{\rm r} \leq N_{\rm t}$. The singular values are then all identical and the singular vectors are Fourier basis vectors, making for very simple precoder and receiver computation.


\begin{figure*}[t]
  \begin{align} 
 h_{n,m} = \exp \! \left(-j \frac{2\pi}{\lambda} \sqrt{ \Bigg( \underbrace{D+n \dr\sin \theta_{\rm r}\cos\phi_{\rm r}-m\dt\sin\theta_{\rm t}}_{z-\text{axis}} \Bigg)^{\!\!2} + \Bigg( \underbrace{n \dr \cos \theta_{\rm r}- m \dt\cos \theta_{\rm t}}_{x-\text{axis}} \Bigg)^{\!\!2} + \Bigg( \underbrace{n \dr \sin \theta_{\rm r} \sin \phi_{\rm r}}_{y-\text{axis}} \Bigg)^{\!\!2} } \, \right) 
  \label{NonparULAs}
  \end{align}
  \hrulefill
\end{figure*}

\begin{figure}
\centering
\subfigure[At the transmitter (respectively the receiver), the clear and shaded circles indicate the location of $m$th transmit antenna (respectively the $n$th receive antenna) and its projection onto the $x$ axis.]{\includegraphics[width =0.999\linewidth]{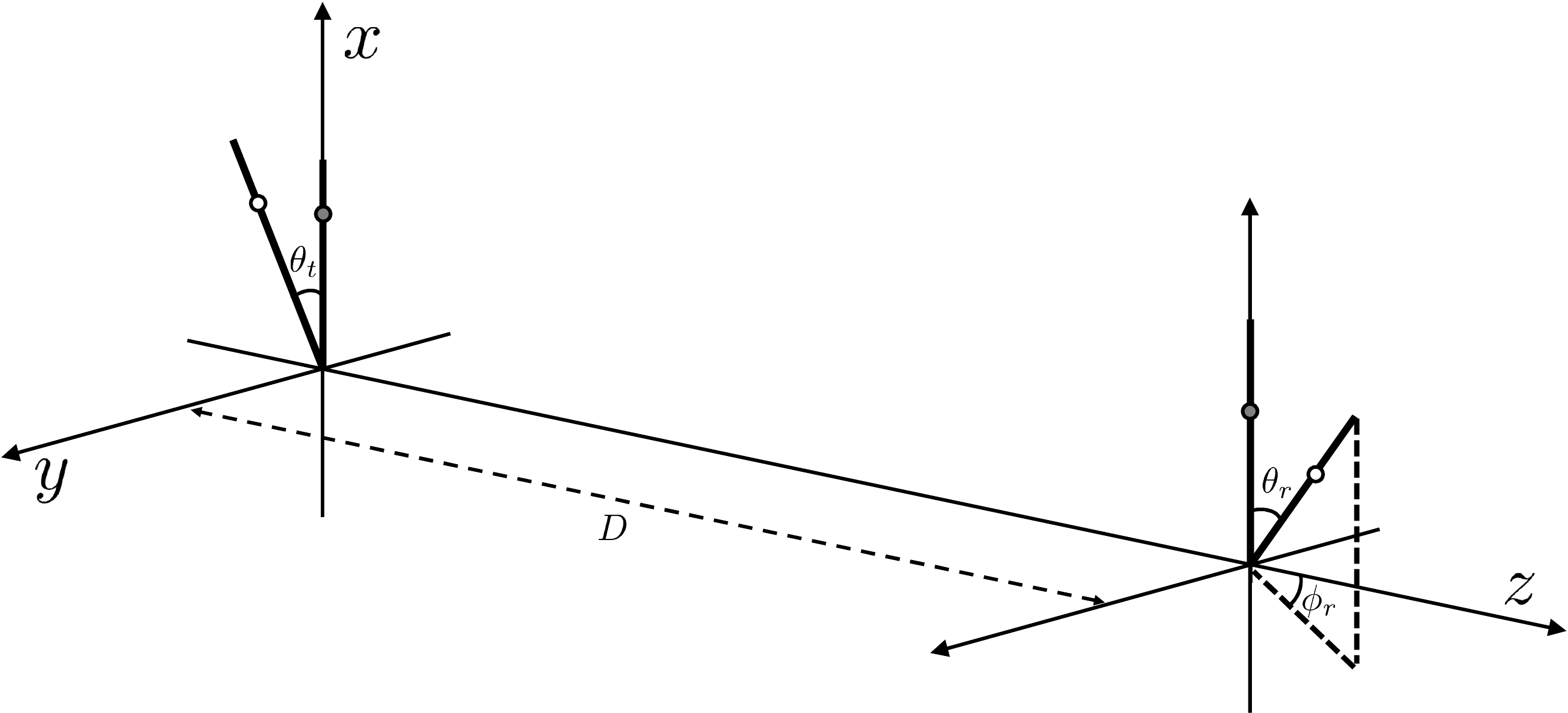}}\\
\subfigure[Projected view on the $xz$ plane.]{\includegraphics[width = 0.999\linewidth]{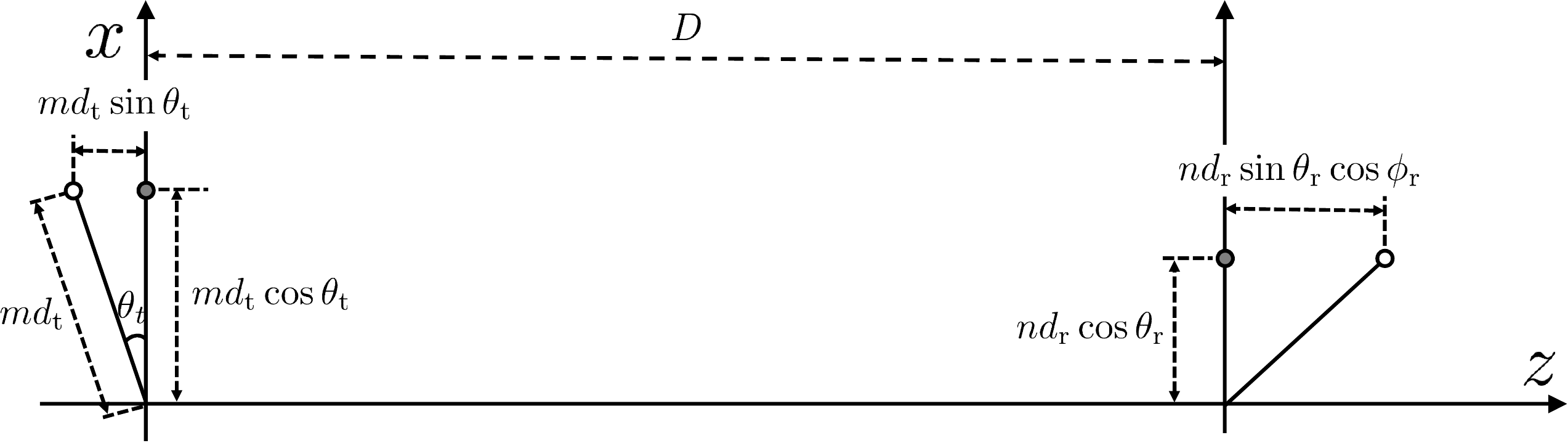}}\\
\subfigure[Projected view on the $yz$ plane.]{\includegraphics[width = 0.999\linewidth]{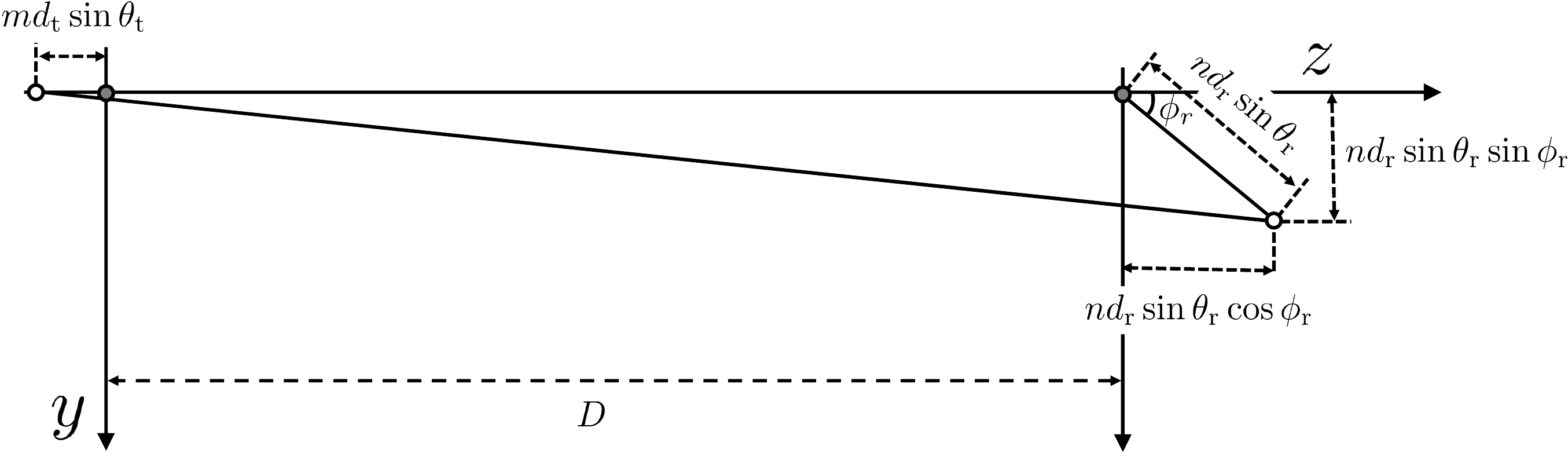}}
\caption{Non-parallel ULAs.}
\label{fig:nonParallelUla}
\end{figure}

\subsection{Non-parallel ULAs}
\label{GranRoyo}

In order to determine in complete generality the relative position of two non-parallel ULAs, three geometric parameters are required, for instance (see Fig. \ref{fig:nonParallelUla}) the elevation angles of the transmit and receive arrays, $\thetat$ and $\thetar$, and the relative azimuth angle between them, $\phir$.
Parameterized by these angles, $h_{n,m}$ is given by (\ref{NonparULAs}) at the top of the next page and, invoking again the premise that the apertures are small relative to $D$, it satisfies
\begin{align}
    h_{n,m}
    & \approx e^{-j2\pi \frac{D}{\lambda}}  \;  e^{-j \pi \left[ \frac{2n}{\lambda} \dr \sin \! \thetar \cos \! \phir + \frac{n^2}{\lambda D} d^2_{\rm r} \, (1-\sin^2 \! \thetar \cos^2 \! \phir) \right] } \nonumber\\
  & \quad \cdot e^{j 2\pi \frac{nm}{\lambda D} \dr \dt \cos \! \thetar \cos \! \thetat} \;  
  e^{-j \pi \left[ \frac{2m}{\lambda} \dt \sin \! \thetat + \frac{m^2 }{\lambda D} d^2_{\rm t} \right] }. \label{eq:nonParallelUlaChannel}
\end{align}
The fixed phase and the phase shifts affecting only either the transmitter or the receiver are again immaterial to the singular values, hence the relevant term is the phase
\be
2\pi \frac{nm}{\lambda D} \dr \dt \cos \thetar \cos \thetat .
\ee
The ensuing normalized channel matrix exhibits the same singular values as $\bH_{\sf ULA}$ in \eqref{Messi}, only with 
\be
\eta=\frac{ (\dr \cos \theta_r) (\dt  \cos \theta_t) \Nmax}{\lambda D}. \label{eq:nonPa_eta}
\ee
Under this more general definition of $\eta$, then, any ULA-induced channel can be represented by $\bH_{\sf ULA}$ in \eqref{Messi} as far as the singular values are concerned. Besides, interestingly, the relative azimuth orientations represented by $\phir$ play no role in the singular values, and therefore in the capacity. This is capitalized on later in the paper.

\section{Capacity Upper Bound}
\label{Iniesta}

We commence by establishing an upper bound on the LOS capacity over all possible antenna placements, a technical result that, besides serving as a benchmark in the sequel, has broad relevance. As detailed in Appendix \ref{Melero}, such upper bound corresponds to a channel having $\rho\in [\Nmin]$ identical nonzero singular values and $\Nmin - \rho$ zero singular values with $\rho$ depending on the SNR via
\begin{align}
    \rho(\SNR) &=
    \begin{cases}
    1  & \quad \qquad\;\:\; \SNR < \zeta_1\\
    2  & \quad \;\; \zeta_1 \leq \SNR < \zeta_2 \\
    3  & \quad \;\; \zeta_2 \leq \SNR < \zeta_3 \\
    \, \vdots \\
    \Nmin  & \quad \!\!\!\! \zeta_{\Nmin-1} \leq \SNR 
    \label{eq:snr_thre}
    \end{cases}
\end{align}
where $\zeta_n$ is a threshold equal to the unique positive solution of
\begin{align}
    f \! \left(\frac{N_{\rm r}N_{\rm t}}{n^2} \, \zeta_n \right)=f \! \left( \frac{N_{\rm r}N_{\rm t}}{(n+1)^2} \, \zeta_n \right)
    \label{CMundo}
\end{align}
given the function
\begin{align}
    f(x)= \frac{1}{\sqrt{x}} \log_2 (1+x ). \label{eq:function}
\end{align}
We then have that
\begin{align}
      C(\SNR) \leq \rho(\SNR) \, \log_2 \! \left(1+ \frac{N_{\rm r}N_{\rm t}}{\rho(\SNR)^2} \SNR \right)
      \label{DeJong}
\end{align}
where the right-hand side is the capacity of such channel with $\rho(\SNR)$ identical nonzero singular values. From (\ref{eq:LoSprop}), these nonzero singular values equal $\sqrt{\Nt\Nr / \rho(\SNR)}$.

By relaxing the integer $\rho$ into a real-valued parameter $\tilde{\rho}\in \mathbb{R}$, a slightly looser upper bound can be obtained in explicit form, precisely
\begin{align}
\label{Xavi}
    C(\SNR) \leq \tilde{\rho}(\SNR) \, \log_2 \! \left(1+ \frac{N_{\rm r}N_{\rm t}}{\tilde{\rho}(\SNR)^2} \SNR \right)
\end{align}
where
\begin{align}
\label{MiriamN}
    \tilde{\rho}(\SNR) = \begin{cases}
    1 & \qquad\qquad \quad \SNR< \frac{c}{\Nmin\Nmax}\\
    \sqrt{\Nmin\Nmax \frac{\SNR}{c}}  & \;  \frac{c}{\Nmin\Nmax} \leq \SNR < \frac{\Nmin c}{\Nmax} \\
    \Nmin & \quad\;\; \frac{\Nmin c}{\Nmax} \leq \SNR 
    \end{cases}
\end{align}
with
\be
c = -1 - \frac{2} { W_0(-2/e^2) }  \approx 3.92 ,
\ee
given $W_0(\cdot)$ as the principal branch of a Lambert W function.
Within the precision of this slightly relaxed bound, then,  the transition away from pure beamforming ($\tilde{\rho}=1$) takes place when $\Nmin\Nmax \SNR$, which is the effective SNR including the beamforming gains, equals precisely $c$.
Combining (\ref{Xavi}) and (\ref{MiriamN}), we can express the relaxed upper bound in the more compact form
\begin{align}
    C(\SNR) \leq \! \begin{cases}
     \log_2 \! \left(1+ N_{\rm r}N_{\rm t} \, \SNR \right) & \qquad \quad\;\; \SNR< \frac{c}{\Nt\Nr}\\
     \! \sqrt{\Nt \Nr \frac{\SNR}{c}}  \log_2(1+c)  &  \frac{c}{\Nt\Nr} \leq \SNR < \frac{\Nmin c}{\Nmax} \\
    \Nmin \log_2 \! \left( \frac{\Nmax}{\Nmin} \SNR \right) & \, \frac{\Nmin c}{\Nmax} \leq \SNR .
    \end{cases}
\end{align}

\begin{figure}
  \centering
  \includegraphics[width=0.93\linewidth]{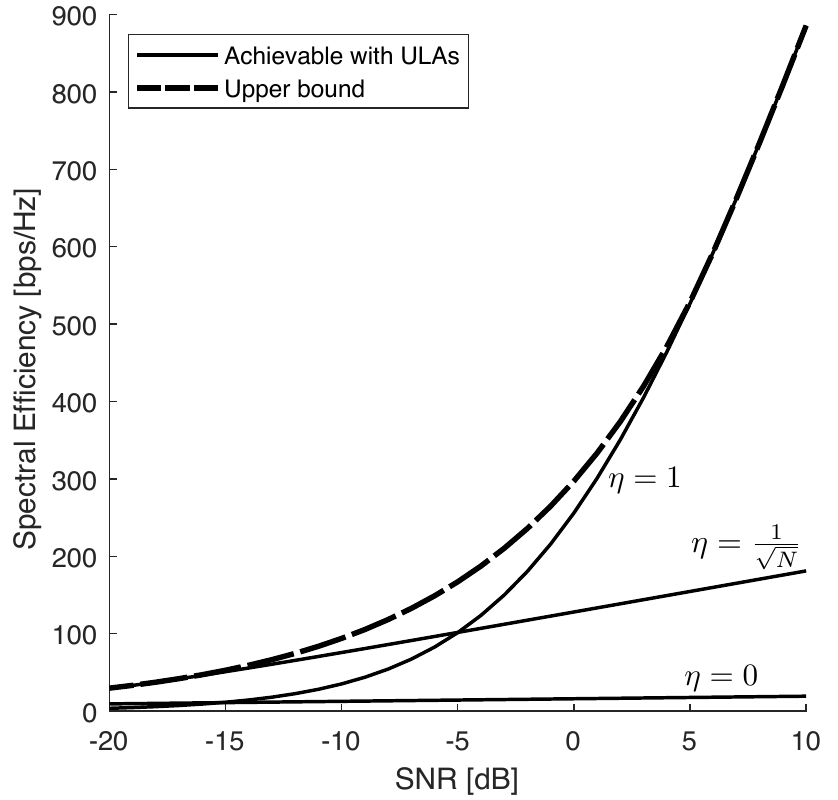}
  \caption{Spectral efficiencies of ULAs with $\eta=0$, $1/\sqrt{N}$, and $1$, for $\Nt=\Nr=N=256$. Also shown is the capacity upper bound.}
  \label{FigULAs}
\end{figure}

 \begin{figure}
  \centering
  \includegraphics[width=0.94\linewidth]{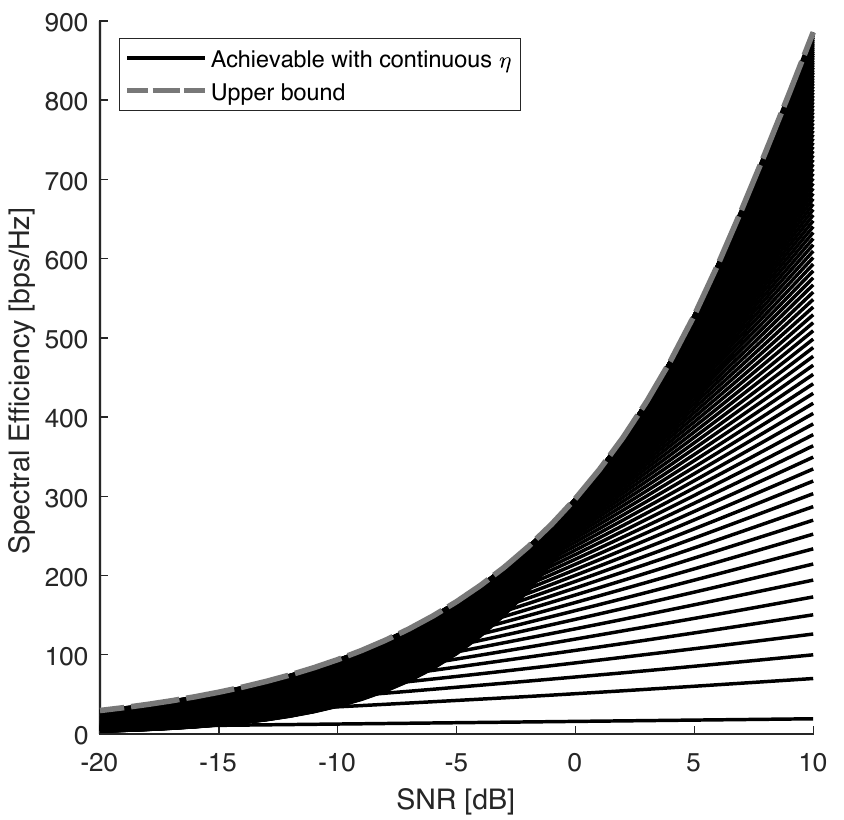}
  \caption{Spectral efficiencies of ULAs with $\eta \in [0,1]$ for $N_{\rm r}=N_{\rm t}=256$. Also shown is the capacity upper bound.}
  \label{FigULAs_2}
\end{figure}

\section{Optimum Antenna Spacings for Parallel ULAs}
\label{ITA}

Parallel ULAs adopting three SNR-based configurations are proposed in \cite{Matthaiou:10}, namely $\eta=0$ for low SNRs, $\eta=1/\sqrt{N}$ for medium SNRs, and $\eta=1$ for high SNRs. Shown in Fig. \ref{FigULAs} are the spectral efficiencies achieved by these configurations for $\Nt=\Nr=N=256$, computed via (\ref{TerStegen}) and (\ref{Messi}), alongside the capacity upper bound. The approach is seen to be effective at very low and at high SNR, less so at intermediate SNRs. At $\SNR=-5$ dB, for instance, the spectral efficiency reaches only about $55\%$ of the capacity.

By releasing $\eta$ and allowing it to take any value within $[0,1]$, the upper bound can be hugged much more closely (see Fig. \ref{FigULAs_2}). This involves fine-tuning the antenna spacings depending on the SNR, computing the singular-value decomposition of $\bH_{\sf ULA}$ to obtain the precoding directions and the receiver, and solving (\ref{TerStegen}) for the transmit powers. 

To interpret the effectiveness of ULAs with antenna spacings adapted to the SNR, let us examine the $(n,m)$th entry of $\bH_{\sf ULA}^*\bH_{\sf ULA}$, namely
\begin{align}
\!\!\!\!    \Big[ \bH_{\sf ULA}^*\bH_{\sf ULA} \Big]_{n,m} 
\!\!\!    &= \!  \sum_{\ell=0}^{N_{\rm r}-1} e^{-j2\pi\eta \frac{(n-m)\ell}{\Nmax}} \\
   & =  \frac{\sin\! \left(\pi\eta \frac{(n-m)N_{\rm r}}{\Nmax} \! \right) \!\!}{\sin\! \left(\pi\eta \frac{n-m}{\Nmax}\right)} \, e^{-j\pi\eta \frac{(n-m)(N_{\rm r}-1)}{\Nmax}} ,\label{eq:dirichlet}
\end{align}
and let us denote by $\lambda_0,\ldots,\lambda_{\Nmin-1}$ the $\Nmin$ largest eigenvalues of  $\frac{\eta}{\Nmax}\bH_{\sf ULA}^*\bH_{\sf ULA}$. For any $\epsilon > 0$, as both  $\Nt$ and $\Nr$ grow large with some ratio $\Nt / \Nr >0$, these eigenvalues can be shown to satisfy \cite{zhu2017eigenvalue,edelman1998future}
\begin{align}
    \lim_{\Nmin\rightarrow \infty}\frac{\big|\big\{ \ell \mid \lambda_\ell < \epsilon \big\}\big|}{\Nmin}=1-\eta
    \end{align}
and    
    \begin{align}
    \lim_{\Nmin\rightarrow \infty}\frac{\big|\big\{ \ell \mid  1-\epsilon<\lambda_\ell <1+\epsilon \big\}\big|}{\Nmin}=\eta
    \label{eq:clusteredEigenvalue}
\end{align}
where $|\{ \cdot  \}|$ indicates the cardinality of a set. These eigenvalues polarize (asymptotically in the numbers of antennas) into the states $1$ and $0$, and therefore the singular values of $\bH_{\sf ULA}$ polarize into $\sqrt{\Nmax/\eta}$ and $0$.
It follows from (\ref{eq:LoSprop}) that (asymptotically) we have $\eta \Nmin$ singular values equal to $\sqrt{\Nmax/\eta}$ and $(1-\eta) \Nmin$ singular values equal to $0$.
By adjusting the antenna spacings such that $\eta=\rho(\SNR)/\Nmin$, we asymptotically obtain a channel matrix
featuring $\rho(\SNR)$ identical nonzero singular values and $\Nmin-\rho$ zero singular values.
This is the precise disposition that yields the capacity upper bound, with $\rho(\SNR)$ as in (\ref{eq:snr_thre}).


Making matters precise, it is shown in Appendix \ref{Monet} that the capacity of the channel $\bH_{\sf ULA}$ with $\eta$ properly adjusted converges pointwise to the LOS capacity, i.e.,
\be
\lim_{\Nmin\rightarrow \infty}\frac{\max_{\eta\in[0,1]} C(\bH_{\sf ULA},\SNR)}{C(\SNR)}=1 
\label{eq:asymptotoicAchievability}
\ee
at every SNR.

For finite numbers of antennas, the polarization of the singular values is not complete and thus the LOS capacity cannot be strictly attained by ULAs in general, but, as illustrated in Fig. \ref{FigULAs_2}, it is approached very closely.
Moreover, there are specific regimes where ULAs can achieve the LOS capacity regardless of the numbers of antennas, as seen next.

\subsection{Low-SNR Regime}
\label{Betis1}

As the SNR drops, $\rho(\SNR)$ shrinks and, for $\SNR \leq \zeta_1=8/\Nr\Nt$,
we have that $\rho(\SNR)=1$. Then, ULAs with tightly spaced antennas become capacity-achieving irrespective of the values of $\Nt$ and $\Nr$, as the channel they create indeed exhibits a single nonzero singular value. Such low-SNR capacity expands as
\begin{align}
C(\SNR) & =  \log_2 \! \left(1 + \Nmin\Nmax \, \SNR \right) \\
& = \Nmin\Nmax \, \SNR \log_2 e + \cO(\SNR^2) .
\end{align}
In contrast, Rayleigh-spaced ULAs would feature $\rho(\SNR)=\Nmin$ and thus 
\begin{align}
C(\bH_{\sf ULA},\SNR) & = \Nmin \log_2\left(1+\frac{\Nmax}{\Nmin}\SNR\right) \\
& =  \Nmax \, \SNR \log_2 e + \cO(\SNR^2) .
\end{align}
With tight ULAs, an $\Nmin$-fold improvement is obtained in low-SNR capacity relative to Rayleigh-spaced ULAs designed for high-SNR operation.


\subsection{High-SNR Regime}
\label{Betis2}

For $\SNR \geq \zeta_{\Nmin-1} $, the upper bound reduces to the capacity of Rayleigh-spaced ULAs, which in this regime become optimum irrespective of the values of $\Nt$ and $\Nr$. Such high-SNR capacity expands as
\begin{align}
C(\SNR) & =  \Nmin \log_2 \! \left(1+\frac{\Nmax}{\Nmin}\SNR\right) \\
& = \Nmin \log_2 \SNR + \cO(1) .
\end{align}

\section{Reconfigurable ULAs}
\label{Netflix}

Adapting $\eta$ to the SNR by means of adjusting the antenna spacings in parallel ULAs is implementationally very challenging because separate moving parts would be required for each individual antenna. This justifies the proposition of having a few arrays with distinct but fixed spacings, and the possibility of switching among them, with a performance shortfall that depends on the number of such arrays \cite{Matthaiou:10}.

This section presents an alternative method to tune $\eta$ as a function of the SNR with a single fixed-spacing ULA at each end of the link.

\subsection{Adapting $\eta$ via ULA Rotation}

To operate with fixed antenna spacings, we seek to adapt $\eta$ by modifying the relative orientation of the ULAs as a function of the SNR. This can be realized by rotating one of the two ULAs, either transmitter or receiver, while keeping the other one fixed. Since, as seen in Section \ref{GranRoyo}, the relative azimuth angle is immaterial to the channel singular values, we set it to $\phir=\pi/2$ (which seems preferable from the standpoint of the directivity of the individual antennas). For starters, we further set $\thetat=0$ and consider rotating only the receiver ULA using $\thetar$. Then, \eqref{eq:nonPa_eta} reduces to
\be
\eta=\frac{(\dr \cos \thetar) \,\dt \Nmax }{\lambda D}  \label{eq:rotation_eta}
\ee
and, by fixing the antenna spacings at the Rayleigh values, further to
\be
\eta=\cos \thetar , \label{eq:rotation_eta2}
\ee
which indeed can be controlled by merely varying $\thetar$. 

Recalling (\ref{eq:asymptotoicAchievability}), we can therefore infer that any two ULAs can achieve capacity (asymptotically in the numbers of antennas) by setting $\phir=\pi/2$, $\thetat=0$, $\dt \dr = \lambda D / \Nmax$, and
\be
\thetar = \arccos  \frac{\rho(\SNR) }{\Nmin} .
\label{GRufian}
\ee
Since $\rho(\SNR) \in [1,\Nmin]$, every SNR maps to a feasible $\thetat$.


If the transmit elevation angle is not set to $\thetat=0$, but rather to some arbitrary value, then (\ref{GRufian}) generalizes to
\be
\thetar = \arccos  \frac{\rho(\SNR) }{\Nmin \cos \thetat} .
\ee


\subsection{Radial ULAs with Electronic Selection}

\begin{figure*}
  \centering
  \includegraphics[width=0.85\linewidth]{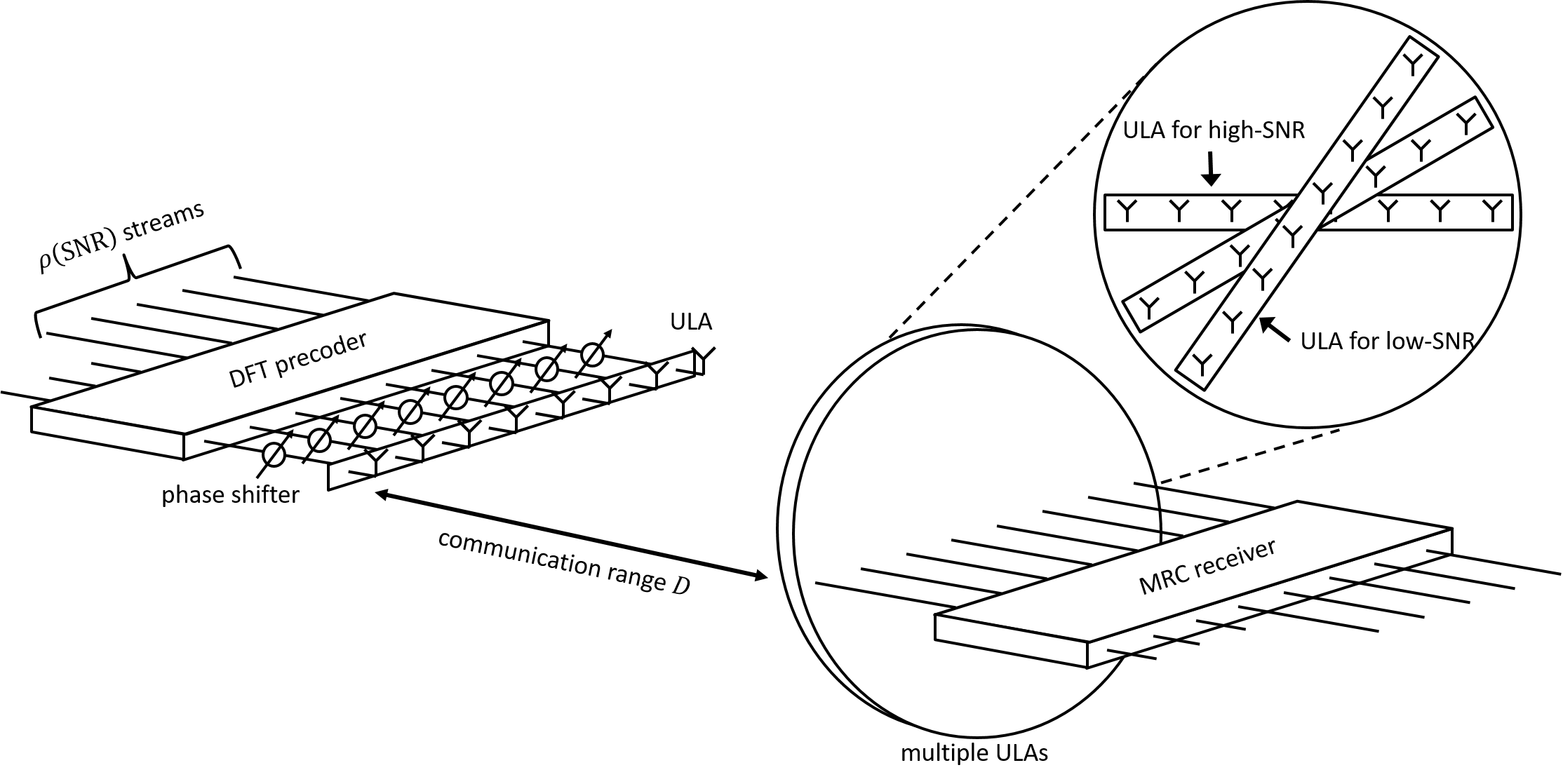}
  \caption{Reduced-complexity configurable architecture.}
  \label{fig:proposedMethod}
\end{figure*}

Since $\rho(\SNR)$ takes on integer values, namely $1,\ldots,\Nmin$, the angles required as per (\ref{GRufian}) to reconfigure a rotating ULA for every possible SNR are discrete. This suggests that, as an alternative to the SNR-based mechanical rotation of the receiver, complete reconfigurability is also possible by deploying at the receiver $\Nmin$ radial ULAs angled at
\be
\theta_{\mathrm{r},n} = \arccos \frac{n}{\Nmin} \qquad\quad n = 1,\ldots,\Nmin .
\ee
Selecting the adequately angled ULA as a function of the SNR (see Fig. \ref{fig:proposedMethod}), we can match the performance of a rotating ULA. The number of radio-frequency chains continues to be $\Nr$, but $\Nmin (\Nr -1)+1$ antennas are now required---the central antenna is common to all the radially arranged ULAs---as opposed to $\Nr$ in the case of a single rotating ULA.
To keep the number of additional antennas to a minimum, the radial architecture should be deployed at the end of the link (transmitter or receiver) with the smallest number of antennas.

Now, a natural next step, especially with a view to having large arrays, is to reduce the number of radial ULAs below $\Nmin$.
%
%
%
Suppose that the values of $\eta$ are taken from the geometric series $1,r,r^2,\ldots$
where the ratio $r<1$ is a design parameter that determines the trade-off between number of ULAs and performance.
To span the entire range $\eta \in [1/\Nmin , 1]$, the number of ULAs required for a given $r$ is
\be
\label{Gloria}
k = 1 + \left \lfloor \frac{ \log \Nmin }{ \log 1/r } \right \rfloor 
\ee
where $\lfloor \cdot \rfloor$ rounds down to the closest integer.
We know that, for a given $\eta$, the channel exhibits (asymptotically)
$\eta \Nmin$ singular values equal to $\sqrt{\Nmax/\eta}$ for a spectral efficiency of
\be
 \eta \Nmin \log_2 \! \left(1+\frac{\Nmax}{\eta^2 \Nmin}\SNR\right) \label{eq:etaCapacity} 
\ee
meaning that, with $k$ radial ULAs, we can hope for a spectral efficiency of
\begin{align}
    R(\SNR) \approx \!\! \max_{\eta \in \{1,r,\ldots, r^{k-1} \}} \eta \Nmin \log_2 \! \left(1+\frac{\Nmax}{\eta^2 \Nmin}\SNR\right)
\end{align}
where the approximation sharpens with the numbers of antennas and the maximization determines the best possible ULA at each SNR.
If we make $\eta$ dependent on the SNR via
\begin{align}
    \eta = \begin{cases}
    r^{k-1} & \qquad\qquad\qquad\qquad\;\; \SNR \leq  \frac{\Nmin}{\Nmax}c r^{2k-3} \\
    \vdots &  \qquad\qquad\qquad\qquad\qquad\quad \vdots \\
     r^2 & \qquad\qquad \frac{\Nmin}{\Nmax}c \, r^5< \SNR \leq \frac{\Nmin}{\Nmax}c \, r^3\\
         r & \qquad\qquad \frac{\Nmin}{\Nmax}c \, r^3< \SNR \leq \frac{\Nmin}{\Nmax}c \, r\\
    1 & \qquad\qquad\;\, \frac{\Nmin}{\Nmax}c \, r < \SNR
    \end{cases} \label{eq:etaGeometric}
\end{align}
where, recall, $c \approx 3.92$, then (see Appendix \ref{Rufian}) the achievable spectral efficiency satisfies
\be
\label{Setien}
\frac{R(\SNR)}{C(\SNR)} \geq \frac{\log(1+c \, r)}{\sqrt{r}\log(1+c)} 
\ee
at every SNR. As $r$ approaches unity, the capacity is approached ever more closely, albeit at the expense of a larger number of radial ULAs as stipulated by (\ref{Gloria}).

If, rather than $\SNR \geq 0$ strictly, we have that $\SNR \geq \SNR_{\sf min} $ with $\SNR_{\sf min} > 0$,
we can truncate \eqref{eq:etaGeometric} into
\begin{align}
    \eta = \begin{cases}
    r^{k-1}& \qquad \frac{\Nmin}{\Nmax}c \, r^{2k-1}< \SNR \leq \frac{\Nmin}{\Nmax}c \, r^{2k-3}\\
    \vdots &  \qquad\qquad\qquad\qquad\qquad\;\; \vdots \\
    r^2 & \qquad\quad\; \frac{\Nmin}{\Nmax}c \, r^5< \SNR \leq \frac{\Nmin}{\Nmax}c \, r^3\\
    r & \qquad\quad\; \frac{\Nmin}{\Nmax}c \, r^3< \SNR \leq \frac{\Nmin}{\Nmax}c \, r\\
    1 & \qquad\quad\;\,\; \frac{\Nmin}{\Nmax}c \, r < \SNR
    \end{cases}
\end{align}
where $k$ now satisfies 
\begin{align}
    \frac{\Nmin}{\Nmax}c \, r^{2k-1}< \SNR_{\sf min} \leq \frac{\Nmin}{\Nmax}c \, r^{2k-3},
\end{align}
ensuring that the SNR range $[\SNR_{\sf min},\infty)$ is covered.
Equivalently,
\begin{align}
\label{Horner}
    k =\left\lfloor \frac{\log \frac{\Nmin \, c}{\Nmax \SNR_{\sf min}}}{2\log 1/r}+\frac{3}{2}\right\rfloor.
\end{align}
This truncation markedly reduces  the number of necessary ULAs, eliminating those that would correspond to SNRs below the operating range of interest.
For instance, with $k = 3$ radial ULAs, $\Nt=\Nr=N$, and $\SNR_{\sf min}=-10$ dB, the largest possible ratio $r$ is, applying (\ref{Horner}), $r=0.48$.
Plugging this value into (\ref{Setien}), we infer that at least $95.9\%$ of the LOS capacity can be achieved (for large $N$) with only three ULAs angled at
$\theta_{\mathrm{r},0}=0$, $\theta_{\mathrm{r},1} = 61^\circ$, and $\theta_{\mathrm{r},2} = 77^\circ$.

\section{Reduced-Complexity Architecture}
\label{sec:complexity}

For $\eta=1$, meaning with Rayleigh antenna spacings at the ULAs, the precoder and the receiver are straightforward to compute as the channel adopts a Fourier structure. This is the situation at high enough SNR, when Rayleigh antenna spacings are optimum. More generally, though, $\eta < 1$; then, obtaining the precoder and receiver requires subjecting the channel matrix to a singular-value decomposition (SVD), with a computational cost of $\mathcal{O} \! \left(\Nmin^2\Nmax\right)$. This is not a limitation in applications where the precoder and receiver recomputation is sporadic, but in more dynamic settings, especially with large arrays, less complex alternatives might be welcome.
In this section, we present one such alternative that very closely approaches the LOS capacity with a computational cost of $\mathcal{O} \!\left(\Nmax\log \Nmax\right)$.

While, for the purpose of capacity calculations, only the singular values of the channel were relevant henceforth, to devise the precoder and the receiver the singular vectors become equally relevant. Then, (\ref{Messi}) no longer suffices to represent nonparallel transmit and receive ULAs, but rather we need to generalize $\bH_{\sf ULA}$ to have its $(n,m)$th entry be given by (\ref{NonparULAs}).
For this more general form of $\bH_{\sf ULA}$, it is shown in Appendix \ref{eleccions} that
\begin{align}
    \bF^*  \bD_{\sf tx}^*\bH^*_{\sf ULA}  \bH_{\sf ULA}  \bD_{\sf tx}  \bF 
    &\approx \frac{\Nmax}{\eta} \, \text{diag} \big(\underbrace{1,\ldots,1}_{\eta \Nmin},0,\ldots,0 \big) \label{eq:nearCirculant}
\end{align}
where $\bD_{\sf tx}$ is a diagonal matrix with entries
\begin{align}
[\bD_{\sf tx}]_{m,m} & = e^{j \pi \left[ \frac{2m}{\lambda} \dt \sin \! \thetat + \frac{m^2 }{\lambda D} d^2_{\rm t} \right] } 
\end{align}
while $\bF$ is a Fourier matrix. Furthermore, the approximation in (\ref{eq:nearCirculant}) tightens as the dimensionality grows large.

\begin{figure}
  \centering
  \includegraphics[width=0.95\linewidth]{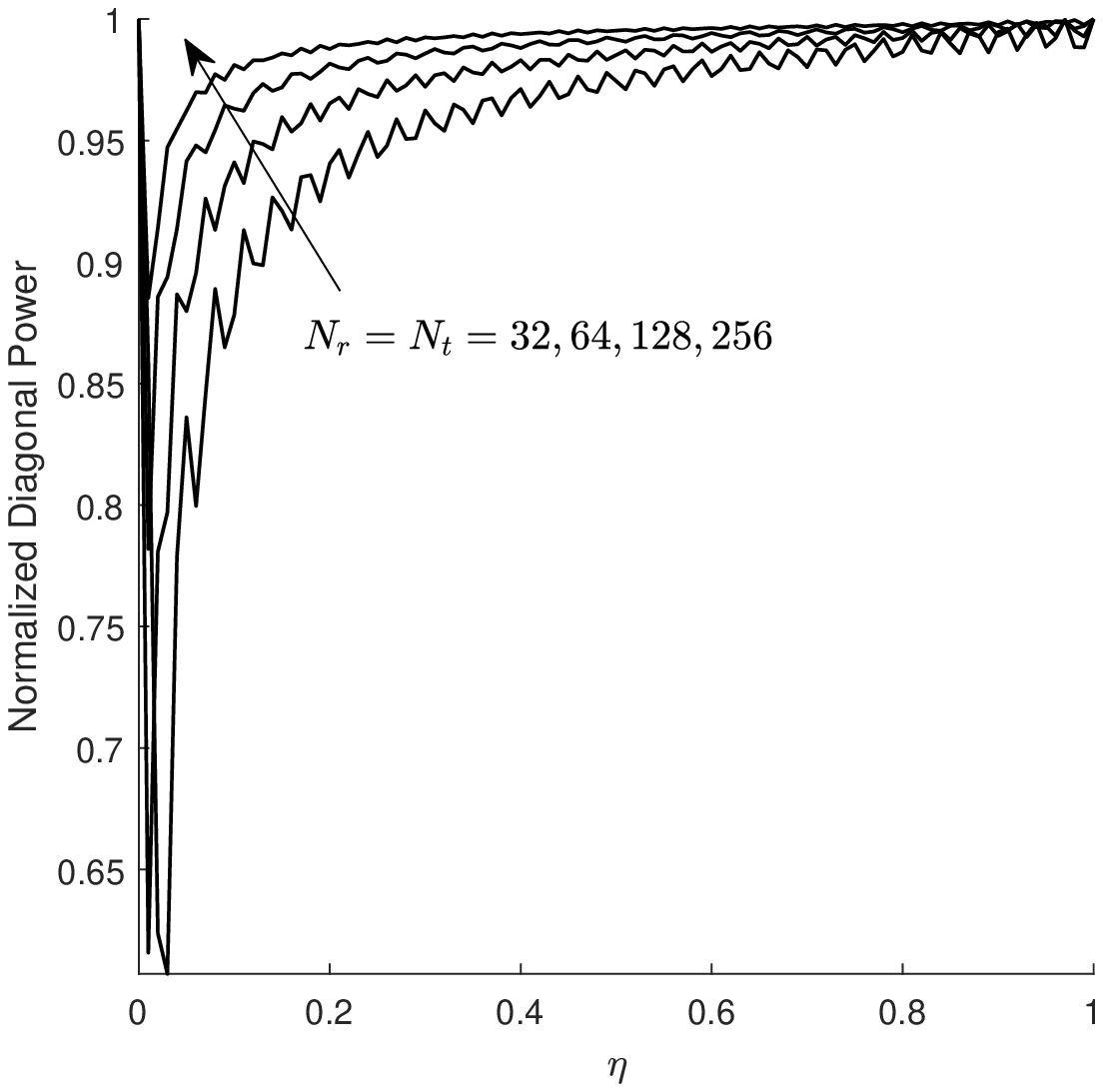}
  \caption{Diagonal power of $\bF^*  \bD_{\sf tx}^*\bH^*_{\sf ULA} \bH_{\sf ULA} \bD_{\sf tx}  \bF$ normalized by its total power for various number of antennas.}
  \label{fig:diagonalPower}
\end{figure}

Asymptotically then, $ \bD_{\sf tx}^*\bH^*_{\sf ULA} \bH_{\sf ULA} \bD_{\sf tx}$ is diagonalized by a Fourier matrix.
This is illustrated in Fig. \ref{fig:diagonalPower}, which depicts the power concentrated on the ensuing diagonal entries relative to the total power, i.e.,
\be
\frac{ \sum_{n=0}^{\Nt-1}[\bF^*  \bD_{\sf tx}^*\bH^*_{\sf ULA} \bH_{\sf ULA} \bD_{\sf tx}  \bF]_{n,n}^2 }{ \| \bF^*  \bD_{\sf tx}^*\bH^*_{\sf ULA} \bH_{\sf ULA} \bD_{\sf tx}  \bF\|_{\rm F}^2 }.
 \ee
Let us see how to take advantage of this behavior.

Inserting between the precoder and the antennas a bank of phase shifts corresponding to the diagonal entries of $\bD_{\sf tx}$, a Fourier precoder $\bF$ yields at the receiver
\be
    \by = \bH_{\sf ULA} \bD_{\sf tx} \bF \bx+\bv .
\ee
With straight maximum ratio combining (MRC) at the receiver, then,
\begin{align}
(\bH_{\sf ULA} \bD_{\sf tx}\bF)^*  \by &= \bF^* \bD^*_{\sf tx} \bH^*_{\sf ULA} \bH_{\sf ULA} \bD_{\sf tx} \bF \bx+\tilde{\bv} 
\end{align}
where the noise at the output of the MRC is
\begin{align}
\tilde{\bv} & = (\bH_{\sf ULA} \bD_{\sf tx}\bF)^*\bv \\
& \sim \mathcal{N}_{\mathbb{C}} \big({\bf 0},\bF^* \bD^*_{\sf tx} \bH^*_{\sf ULA} \bH_{\sf ULA} \bD_{\sf tx} \bF \big).
\end{align}
From \eqref{eq:nearCirculant}, this transmit-receive architecture (asymptotically) decomposes the channel into $\eta \Nmin$ parallel subchannels, each with power gain $(\Nmax/\eta)^2$ and noise variance $\Nmax/\eta$. By transmitting $\eta \Nmin$ equal-power signal streams over these subchannels and separately decoding them, the spectral efficiency (asymptotically) equals \eqref{eq:etaCapacity}.


Applying the receiver matrix $(\bH_{\sf ULA} \bD_{\sf tx}\bF)^*$ to the vector $\by$ entails a complexity of $\cO(\Nr\Nt) = \cO(\Nmin \Nmax)$, already markedly lower than the $\cO(N^2_{\sf min} \Nmax)$ that an SVD necessitates. A further reduction can be attained by
rearranging the receiver into
\begin{align}
    (\bH_{\sf ULA} \bD_{\sf tx}\bF)^*=\bF^*(\bD_{\sf rx}\bH_{\sf ULA} \bD_{\sf tx})^*\bD_{\sf rx}
\end{align}
where $\bD_{\sf rx}$ is a diagonal matrix with entries
\begin{align}
    [\bD_{\sf rx}]_{n,n}=e^{j2\pi \frac{D}{\lambda}} e^{j \pi \left[ \frac{2n}{\lambda} \dr \sin \! \thetar \cos \! \phir + \frac{n^2}{\lambda D} d^2_{\rm r} \, (1-\sin^2 \! \thetar \cos^2 \! \phir) \right] }.
\end{align}
The application of $\bF^*$ and $\bD_{\sf rx}$ to a vector require complexities of $\cO(\Nt\log\Nt)$ and $\cO(\Nr)$, respectively. In turn, $(\bD_{\sf rx}\bH_{\sf ULA} \bD_{\sf tx})^*$ has entries
\begin{align}
    [(\bD_{\sf rx}\bH_{\sf ULA} \bD_{\sf tx})^*]_{n,m}=e^{-j 2\pi\eta \frac{nm}{\Nmax}}
\end{align}
and, from the relationship
\begin{align}
    e^{-j 2\pi\eta \frac{nm}{\Nmax}}=e^{-j \pi\eta \frac{n^2}{\Nmax}}e^{j \pi\eta \frac{(n-m)^2}{\Nmax}}e^{-j \pi\eta \frac{m^2}{\Nmax}},
\end{align}
it follows that $(\bD_{\sf rx}\bH_{\sf ULA} \bD_{\sf tx})^*$ is a Toeplitz matrix pre- and post-multiplied by diagonal matrices. Therefore, the complexity of applying $(\bD_{\sf rx}\bH_{\sf ULA} \bD_{\sf tx})^*$ to a vector is $\cO(\Nmax\log\Nmax)$ \cite{gohberg1994fast}. Altogether, with a precoding complexity of $\cO(N_{\rm t} \log N_{\rm t})$ and a receiver complexity of $\cO(\Nmax\log\Nmax)$, the performance equals (asymptotically) that of an SVD-based architecture.

\begin{figure}
  \centering
  \includegraphics[width=0.94\linewidth]{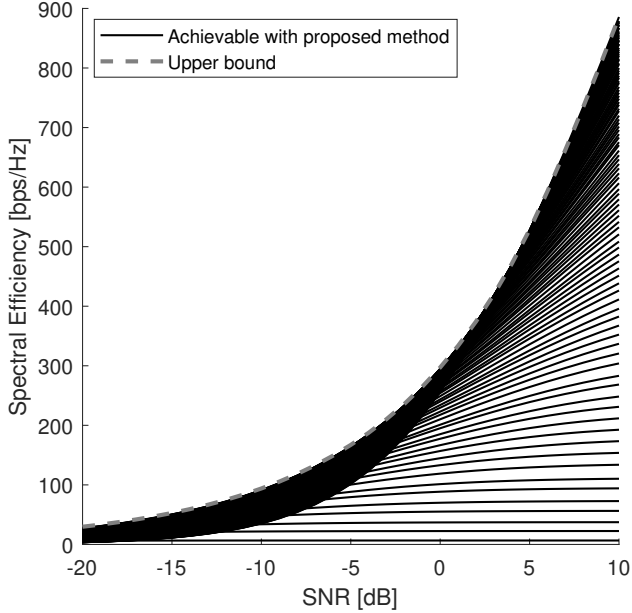}
  \caption{Spectral efficiencies for $\eta \in [1/\Nmin,1]$ achievable by the proposed reduced-complexity architecture with $N_{\rm r}=N_{\rm t}=256$. Also shown, in dashed, is the capacity upper bound.}
  \label{fig:achievableRateWithProposedMethod}
\end{figure}

Fig. \ref{fig:achievableRateWithProposedMethod} illustrates how the reduced-complexity architecture consisting of a Fourier precoder and a bank of phase shifts at the transmitter, and an MRC at the receiver, can very tightly track the LOS capacity upper bound. 
Complementing the figure, Table \ref{table:parameterAndSpectralEfficiency} compares the performance of the low-complexity architecture against those of its SVD-based counterpart and of parallel Rayleigh-spaced ULAs, for various SNRs and numbers of antennas.
For $\SNR=-10$ dB and $N_{\rm r}=N_{\rm t}=256$, for instance, parallel ULAs attain only $37.5\%$  of the capacity upper bound whereas, by applying a rotation of $\thetar=80.8\degree$, that share increases to $ 99.5\%$ (with SVD-based precoder and receiver) or $96.4\%$ (with the lower-complexity architecture).

\subsection{Comparison with Other Architectures}

The presence of the Fourier precoder is very consequential in the proposed reduced-complexity architecture.
Without it, an MRC receiver would perform poorly because, from \eqref{eq:clusteredEigenvalue},
\begin{align}
    \|\bH_{\sf ULA}^*\bH_{\sf ULA}\|_{\rm F}^2 
    &= \!\! \sum_{n=0}^{\Nmin-1} \!\! \lambda_n^2 \\
    &\approx \eta \Nmin \left(\frac{\Nmax}{\eta}\right)^{\!2} \\
    & = \frac{\Nmin\Nmax^2}{\eta}
\end{align}
while the power on the diagonal entries adds up to
\begin{align}
\sum_{n=0}^{\Nmin-1} [\bH_{\sf ULA}^*\bH_{\sf ULA}]_{n,n}^2 
&= \Nmin\Nmax^2    
\end{align}
regardless of $\eta$. Thus, the share of power on the diagonal equals $\eta \leq 1$, meaning that there is more power on the off-diagonal entries (which represent interference) than on the diagonal entries (which represent intended signals).

\begin{table}
\caption{Configuration parameters and spectral efficiency (as a share of the LOS capacity upper bound) with Rayleigh-spaced ULAs and $\Nr=\Nt=N$.}
\label{table:parameterAndSpectralEfficiency}
\begin{center}
\begin{tabular}{|c|cccc|c|}
\hline
SNR [dB]                & -20           & -10            & 0             & 10         & $N$ \\
\hline
$\eta$                  & 0.05          & 0.16          & 0.5           & 1.0       & \text{any} \\
$\theta_{\rm r}$        & 87.1\degree   & 80.8\degree   & 59.7\degree   & 0\degree  & \text{any} \\
\hline
\text{Parallel}         & 12.4\%        & 37.5\%        & 86.1\%        & 100\%     & $256$\\
Rotated (SVD)           & 98.6\%        & 99.5\%        & 99.8\%        & 100\%     & $256$\\
Rotated (Fourier+MRC)   & 89.1\%        & 96.4\%        & 99.1\%        & 100\%     & $256$\\
\hline
\text{Parallel}         & 12.5\%        & 37.5\%        & 86.1\%        & 100\%     & $32$\\
Rotated/SVD             & 95.0\%        & 97.1\%        & 99.1\%        & 100\%     & $32$\\
Rotated (Fourier+MRC)   & 37.5\%        & 84.0\%        & 95.2\%        & 100\%     & $32$\\
\hline
\end{tabular} 
\end{center}
\end{table}

\section{Conclusion}
\label{sec:conclusion}

This paper has shown that, through an SNR-dependent rotation, ULAs with Rayleigh antenna spacings can be configured to closely approach the LOS channel capacity. 
Such capacity is actually attained asymptotically in the numbers of antennas, and at low/high SNR.
The same performance can be achieved, avoiding the need for mechanical rotations, by selecting among $\Nmin$ radially disposed ULAs.
As the number of radial ULAs shrinks, the performance declines very gradually such that, with only a few properly oriented ULAs, the vast majority of the capacity is still within reach.

In comparison with structures consisting of multiple ULAs featuring distinct antenna spacings, the proposed architecture is more compact and better performing. A structure with three distinct-spacing ULAs, for instance, performs very well at some SNRs, but drops down to only $55\%$ of capacity at others. In contrast, the proposed architecture ensures at least $96\%$ of the LOS capacity at every SNR.

Potential follow-up research directions include the extension to a variety of antenna configurations such as uniform rectangular arrays (URAs), which offer superior form factors. It is also interesting to optimize antenna arrangements under uniform circular arrays (UCAs) structure, which has a deep connection with orbital angular momentum multiplexing techniques using multiantenna systems \cite{bai2014experimental, opare2015degrees}.
Another worthwhile direction is the analysis of the robustness in the face of residual multipath propagation, i.e., in channels exhibiting Rice fading \cite[sec. 3.4]{Foundations:18} rather than being purely LOS.


%

\appendices
\section{}
\label{Melero}

A proof of (\ref{DeJong}) can be found in \cite{ISIT}. Here, we provide an alternative proof that relies on the following intuitive result.

\begin{lemma}
\label{lemma1}
Consider $N$ real numbers $x_1,\ldots, x_N$. If we successively replace two of them with their average, it is possible to make all of them to be equal in the limit. Precisely, letting $t$ be an iteration counter, there exists a sequence satisfying
\begin{align}
    \left( x_1^{(t)},\ldots, x_N^{(t)} \right) \stackrel{t \to \infty}{\rightarrow} \left(\frac{\sum_{i=1}^N x_i}{N},\ldots, \frac{\sum_{i=1}^N x_i}{N}\right).
\end{align}
\end{lemma}
\begin{IEEEproof}
Averaging the largest and smallest values does the trick. Since the sum of $N$ numbers is invariant under the averaging operations, we can let $x_1+\cdots+x_N = 0$ without loss of generality. Now, observe $\big(x_1^{(t)}\big)^2+\cdots+\big(x_N^{(t)}\big)^2$ at iteration $t$. After one more iteration, this sum decreases by
\begin{align}
\left(\! \max x_i^{(t)} \!\right)^{\!2}\! + \left(\! \min x_i^{(t)}\! \right)^{\!2}\! - 2 \left(\! \frac{ \max x_i^{(t)}\! + \min x_i^{(t)}  }{2} \!\right)^{\!\!2} \nonumber \\
 =  \frac{\left(\!\max x_i^{(t)} \!- \min x_i^{(t)} \!\right)^{\!2}}{2}.
\end{align}
From the zero sum constraint, $\max x_i^{(t)}\geq 0 \geq \min x_i^{(t)}$, and thus
\begin{align}
\frac{\left(\!\max x_i^{(t)} \!- \min x_i^{(t)} \!\right)^{\!2}}{2} & \geq \frac{\left(\max \big| x_i^{(t)} \big| \right)^{\!2}}{2} \\
& \geq \frac{\left(x_1^{(t)}\right)^2+\cdots+\left(x_{N}^{(t)}\right)^{\!2}}{2N}.
\end{align}
where we capitalized on the fact that the maximum is no smaller than the average.
Altogether,
\begin{align}
&  \!\!\!\! \left(x_1^{(t+1)}\right)^{\!2}+\cdots+\left(x_N^{(t+1)}\right)^{\!2} \nonumber \\
    &  \!\!\!\! \qquad\qquad \leq \left(1 - \frac{1}{2N} \right)  \left[\left(x_1^{(t)}\right)^{\!2}+\cdots+\left(x_N^{(t)}\right)^{\!2}\right]\\
    &  \!\!\!\! \qquad\qquad \leq \left(1 - \frac{1}{2N} \right)^{\! t+1}\left[\left(x_1^{(0)}\right)^2+\cdots+\left(x_N^{(0)}\right)^2\right] \\
&  \!\!\!\! \qquad\qquad     \stackrel{t \to \infty}{\rightarrow} 0.
\end{align}
\end{IEEEproof}

\begin{figure}
  \centering
  \includegraphics[width=0.9\linewidth]{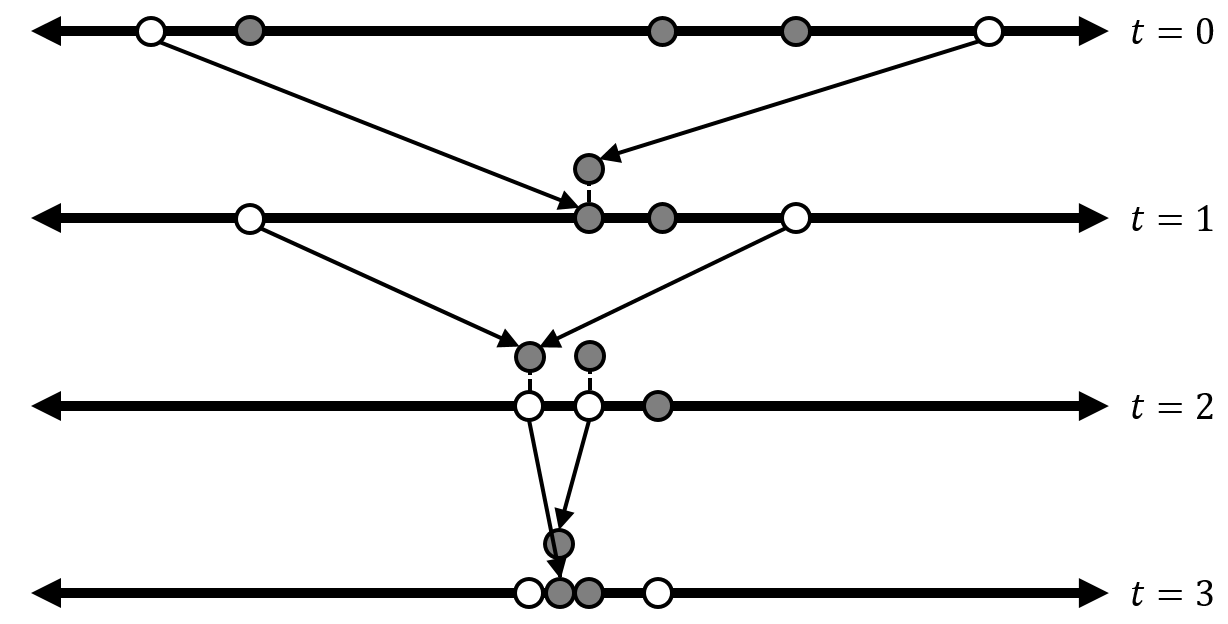}
  \caption{Exemplification of the averaging process in a sequence of iterations. Each point represents a real number $x_i$, with the largest and smallest at each iteration colored in gray. The distribution converges to a mass point as the iterations progress.}
\end{figure}

Combining (\ref{TerStegen}) and (\ref{eq:optFormulation}) we obtain, as starting point,
\begin{align}
 C & = \!\!   \max_{\substack{\sum_{n=0}^{\Nmin-1} \! \sigma_n^2 = \Nr\Nt \\ \sigma_n^2 \geq 0}}\max_{\substack{\sum_{n=0}^{\Nmin-1} \! p_n = {\sf SNR}\\ p_n \geq 0}} \sum_{n=0}^{\Nmin-1} \!\! \log_2 \! \left(1+\sigma_n^2 \, p_n\right) . 
 \label{Pique}
\end{align}
Then, defining
\begin{align}
\bar{\sigma}_n^2 & =  \frac{\sigma_n^2/\Nr\Nt+p_n/{\sf SNR}}{2} \, \Nr\Nt \\
\bar{p}_n & =  \frac{\sigma_n^2/\Nr\Nt+p_n/{\sf SNR}}{2} \, \SNR ,
\end{align}
we have that the argument of (\ref{Pique}) satisfies, by virtue of the inequality between the arithmetic and geometric means,
\begin{align}
   & \!\! \sum_{n=0}^{\Nmin-1} \!\! \log_2\!\left(1 + \sigma_n^2 \, p_n\right) \\ 
   & \qquad\quad  \leq \! \sum_{n=0}^{\Nmin-1} \! \log_2 \! \left(1+  \frac{ \Nr\Nt \, \SNR}{2}  \! \left( \frac{\sigma_n^2}{\Nr\Nt} + \frac{p_n}{\SNR} \right)^{\!2} \right) \nonumber \\
    & \qquad\quad = \sum_{n=0}^{\Nmin-1} \! \log_2 \! \left(1+\bar{\sigma}_n^2 \, \bar{p}_n\right)
\end{align}
under constraints that are preserved, namely
\begin{align}
\sum_{n=0}^{\Nmin-1} \bar{\sigma}_n^2 & = \Nr\Nt \\
 \sum_{n=0}^{\Nmin-1} \bar{p}_n & = \SNR .
\end{align} 
 From the relationship $\bar{\sigma}_n^2=\frac{\Nr\Nt}{{\sf SNR}}\bar{p}_n$, then,
\begin{align}
 C \leq   \max_{\substack{\sum_{n=0}^{\Nmin-1} \bar{p}_n = \SNR \\ \bar{p}_n \geq 0}} \sum_{n=0}^{\Nmin-1} \log_2 \! \left(1+\frac{\Nr\Nt}{\sf SNR} \, \bar{p}_n^2\right) .
\label{eq:onlyPower}
\end{align}

Now, armed with Lemma \ref{lemma1}, we set out to prove by induction that
\begin{align}
   & \max_{\substack{\sum_{n=0}^{\Nmin-1} {p}_n = \SNR \\ {p}_n \geq 0}}  \sum_{n=0}^{\Nmin-1} \!\! \log_2 \! \left(1+\frac{\Nr\Nt}{\sf SNR} \, {p}_n^2\right) \nonumber \\
    & \qquad\qquad = \!\! \max_{\rho \in \{1,2,\ldots,\Nmin\}} \rho \log_2\! \left(1+\frac{\Nr\Nt}{\rho^2}\SNR\right) .
\end{align}
Introducing $x_n=\sqrt{\Nr\Nt/\sf SNR} \, {p}_n$ and $X=\sqrt{\Nr\Nt\SNR}$, the above becomes
\begin{align}
   & \max_{\substack{\sum_{n=0}^{\Nmin-1} x_n = X \\ x_n \geq 0}}  \sum_{n=0}^{\Nmin-1} \!\! \log_2 \! \left(1+x_n^2\right)\\
    & \qquad\qquad = \!\! \max_{\rho \in \{1,2,\ldots,\Nmin\}} \rho \log_2\! \left(1+\frac{X^2}{\rho^2}\right).
    \label{ITA20}
\end{align}
We start from the base case $\Nmin=2$, which boils down to a single variable optimization given that $x_2 = X-x_1$. From
\begin{align}
& \log_2 \! \left(1+x_1^2\right)+\log_2 \! \left(1+(X-x_1)^2\right) \nonumber \\
& \qquad\qquad =\log_2 \! \Big(\!\left(1+x_1^2\right)\left(1+(X-x_1)^2\right)\Big),
\end{align}
it suffices to show that, when the domain is $[0,X]$, the quartic function $\left(1+x_1^2\right)\left(1+(X-x_1)^2\right)$ can attain its maximum only at $0$, $X/2$, or $X$. This quartic function is symmetric along $x_1=X/2$. If the quartic function has its local maximum at $\bar{x}\neq 0,X/2,X$, then it has one more local maximum at $X-\bar{x}$ from symmetry. This contradicts to the fact that quartic function has at most one local maximum, which completes the proof for the base case.

Assuming now that (\ref{ITA20}) holds for $\Nmin=N$, let us prove it for $\Nmin=N+1$. Let $x_0, x_1, \ldots, x_{N-1}$, whose sum is $X$, be given. Now, we replace the maximum and minimum ones with $\bar{x}_1$ and $\bar{x}_2$ maximizing
\begin{align}
\log_2 \! \left(1+\bar{x}_1^2\right)+\log_2 \! \left(1+\bar{x}_2^2\right) \label{eq:maxTwo}
\end{align}
and satisfying
\be
\bar{x}_1+\bar{x}_2 = \max\limits_{i}x_i+\min\limits_{i}x_i.
\ee
 From the base case, $(\bar{x}_1,\bar{x}_2)$ turns out to equal either
\begin{align}
    \left(\max\limits_{i}{x}_i+\min\limits_{i}{x}_i,0\right),\left(0,\max\limits_{i}{x}_i+\min\limits_{i}{x}_i\right) \label{eq:onePower}
\end{align}
or
\begin{align}
\left(\frac{\max\limits_{i}{x}_i+\min\limits_{i}{x}_i}{2},\frac{\max\limits_{i}{x}_i+\min\limits_{i}{x}_i}{2}\right). \label{eq:twoPower}    
\end{align}
Over the successive replacements, if at some iteration $t$ it happens that $(\bar{x}_1^{(t)},\bar{x}_2^{(t)})$ equal \eqref{eq:onePower}, the induction argument becomes complete. Otherwise, with the averaging operation describe in the lemma being performed at each iteration, we know that
\begin{align}
    \sum_{n=0}^{N} \log_2 \! \left(1+(x_n^{(0)})^2\right)
    &\leq \sum_{n=0}^{N}  \log_2 \! \left(1+(x_n^{(t)})^2\right)\\
    &\stackrel{t \to \infty}{\rightarrow} (N+1)\log_2 \! \left(1+\frac{X^2}{(N+1)^2}\right) .
\end{align}


\section{}
\label{Monet}

For $\SNR\geq \frac{\Nmin c}{\Nmax}$, the LOS capacity upper bound is achieved by setting $\eta=1$, hence we concentrate on proving (\ref{eq:asymptotoicAchievability}) for $\SNR<\frac{\Nmin c}{\Nmax}$.
The numerator of (\ref{eq:asymptotoicAchievability}) is lower bounded by any achievable spectral efficiency, and in particular the one obtained by setting
\be
\eta = \bar{\eta}=\sqrt{\frac{\Nmax}{\Nmin}\frac{\SNR}{c}} \label{eq:etaChoice}
\ee
and driving only the subchannels with gain larger then $\Nmax \, (1-\epsilon)/\bar{\eta}$. This gives
\begin{align}
& \!\!\!\!   \max_{\eta \in [0,1]} C(\bH_{\sf ULA},\SNR)  \geq \Big|\Big\{ \ell \mid  1-\epsilon <\lambda_\ell <1+\epsilon \Big\}\Big| \\
& \!\!\!\!    \qquad\quad\; \cdot \log_2 \! \left( \! 1 + \frac{\Nmax \, (1-\epsilon)}{\bar{\eta}}\frac{\SNR}{\big|\big\{ \ell \mid  1-\epsilon\!<\!\lambda_\ell \!<\!1+\epsilon \big\}\big|}\right) \nonumber \\
& \!\!\!\!    \qquad\qquad\qquad\qquad\quad =\Nmin \, \frac{\big|\big\{ \ell \mid  1-\epsilon<\lambda_\ell < 1+\epsilon \big\}\big|}{\Nmin}\nonumber\\
& \!\!\!\!    \qquad\quad\; \cdot \log_2 \!\left( 1 + \frac{\frac{\Nmax}{\Nmin} \, (1-\epsilon)}{\bar{\eta}}\frac{\SNR}{\frac{|\{ \ell \, \mid \,  1-\epsilon<\lambda_\ell <1+\epsilon \}|}{\Nmin}}\right) ,
\end{align}
which, from the asymptotic polarization of the eigenvalues, leads to
\begin{align}
    & \!\!\!\! \lim_{\Nmin \to \infty}  \frac{ \max_{\eta \in [0,1]} C(\bH_{\sf ULA},\SNR)}{\Nmin} \nonumber \\
     & \qquad\qquad\qquad\quad  \geq \bar{\eta}\log_2 \!\left( 1 + (1-\epsilon) \frac{\Nmax}{\Nmin} \frac{\SNR}{\eta^2}\right) \\
    & \qquad\qquad\qquad\quad =\bar{\eta}\log_2 \! \Big( 1+(1-\epsilon) \, c \Big)
\end{align}
where the last step follows from \eqref{eq:etaChoice}.

Turning now to the denominator of (\ref{eq:asymptotoicAchievability}), it is upper bounded---recall Section \ref{Iniesta}---by 
\be
\Nmin\sqrt{\frac{\Nmax}{\Nmin}\frac{\SNR}{c}}\log_2 (1+c )=\Nmin\cdot \bar{\eta} \log_2(1+c).
\ee
Altogether then, (\ref{eq:asymptotoicAchievability}) satisfies
\begin{align}
\label{COV}
    \lim_{\Nmin\rightarrow \infty}\frac{\max_{\eta \in [0,1]}  C(\bH_{\sf ULA},\SNR)}{C(\SNR)}
    &\geq\frac{\log_2 \! \big( 1+(1-\epsilon) \, c \big)}{\log_2(1+c)}
\end{align}
where $\epsilon$ can be arbitrarily small, making the ratio arbitrarily close to $1$.
We note that this pointwise convergence does not imply uniform convergence, and in fact
\be
\lim_{\Nmin\rightarrow \infty}\min_{\SNR}\frac{\max_{\eta\in[0,1]}C(\bH_{\sf ULA},\SNR)}{C(\SNR)} < 1
\ee
because of the behavior around the SNR thresholds identified in Section \ref{Iniesta}.
Notwithstanding that, for the purposes of this paper ULAs asymptotically achieve the LOS capacity.

\section{}
\label{Rufian}

Suppose that 
\be
\frac{\Nmin}{\Nmax}cr^{2\ell+1}<\SNR<\frac{\Nmin}{\Nmax}cr^{2\ell-1},
\ee
such that $\eta$ in \eqref{eq:etaGeometric} equals $r^\ell$ ($\ell=0,1,\ldots,k-1$). In this interval,
\begin{align}
     \frac{R(\SNR)}{C(\SNR)}\geq\frac{r^\ell \Nmin \log_2 \! \left(1+\frac{\Nmax}{\Nmin r^{2\ell}}\SNR\right)}{\sqrt{\Nmin\Nmax} \sqrt{\frac{\SNR}{c}} \log_2(1+c)}.
\end{align}
With the substitution $x=\frac{\Nmax}{\Nmin r^{2\ell}}\SNR$, the right-hand side becomes $f(x)/f(c)$ with the above SNR range mapping to $cr \leq x \leq c/r$ and with $f(\cdot)$ as defined in (\ref{eq:function}).
Since $f(\cdot)$ is unimodal with maximum attained at $c$,
\begin{align}
\min_{cr \leq x \leq c/r} \frac{f(x)}{f(c)} & = \frac{\min \! \big( f(cr),f(c/r) \big)}{f(c)} \\
& = \frac{ f(cr)}{f(c)}
\end{align}
where the last step follows from $f(cr)\leq f(c/r)$, as can be verified.


\section{}
\label{eleccions}

This appendix builds on \cite{wang2014tens} to establish (\ref{eq:nearCirculant}).
Letting $\bT = \bH_{\sf ULA}^*\bH_{\sf ULA}$ with
\begin{align}
    t_{n-m}=e^{-j\pi\eta \frac{(n-m)(N_{\rm r}-1)}{\Nmax}}\frac{\sin\! \left(\pi\eta \frac{(n-m)N_{\rm r}}{\Nmax} \! \right) \!\!}{\sin\! \left(\pi\eta \frac{n-m}{\Nmax}\right)} \, 
\end{align}
its $(n,m)$th entry---the indexing depends only on the difference between $n$ and $m$ because $\bT$ is a Toeplitz matrix---we aim to show that
\begin{align}
    & \Big\|\bF^*\bT\bF - \frac{\Nmax}{\eta} \, \text{diag} \big(\underbrace{1,\ldots,1}_{\floor{\eta \Nmin}},0,\ldots,0 \big)\Big\|_{\rm F}^2 \label{Mobile} \\
    & \qquad\qquad = \Big\|\bT - \frac{\Nmax}{\eta} \, \bF \, \text{diag} \big(\underbrace{1,\ldots,1}_{\floor{\eta \Nmin}},0,\ldots,0 \big)\bF^*\Big\|_{\rm F}^2 \nonumber
\end{align}
is $\mathcal{O}(N^2_{\rm t}\log \Nt)$ whereas $\|\bT\|_{\rm F}^2=\Theta(N^3_{\rm t})$, meaning that the relatively difference between the terms being subtracted in (\ref{Mobile}) diminishes as $\Nr$ and $\Nt$ grow large with a fixed ratio. Defining
\be
\bC = \frac{\Nmax}{\eta} \, \bF \, \text{diag} \big(\underbrace{1,\ldots,1}_{\floor{\eta \Nmin}},0,\ldots,0 \big)\bF^* ,
\ee
the $(n,m)$th entry of $\bC$, also Toeplitz, is given by
\begin{align}
c_{n-m}=e^{-j\pi \frac{(n-m)(\lfloor\eta\Nmin\rfloor-1)}{\Nt}} \frac{\sin\! \left(\pi\frac{(n-m)\lfloor\eta\Nmin\rfloor}{\Nt}\right)}{\frac{\eta \Nt}{\Nmax}\sin\! \left(\pi\frac{n-m}{\Nt}\right)}.
\end{align}
With $\bT$ and $\bC$ being Toeplitz matrices, their squared difference can be written as
\be
\label{FMA}
\|\bT-\bC\|_{\rm F}^2=\sum_{|\ell|<\Nt}(\Nt-|\ell|) \, |t_\ell-c_\ell|^2 .
\ee
By means of the quantities defined as
\begin{align}
    & x_\ell \!=\! \sin\! \frac{\pi\eta\ell}{\Nmax} &&x_\ell + \Delta x_\ell \!=\! \frac{\eta\Nt}{\Nmax}\sin \frac{\pi\ell}{\Nt}\\
    & y_\ell \!=\! \sin\! \frac{\pi\eta\ell\Nr}{\Nmax} &&  y_\ell + \Delta y_\ell \!=\! \sin \frac{\pi\ell\floor{\eta\Nmin}}{\Nt}\\
    & \theta_\ell \!=\! -\pi\eta \frac{\ell(N_{\rm r}-1)}{\Nmax} && \theta_\ell + \Delta \theta_\ell \!=\! -\pi \frac{\ell(\lfloor\eta\Nmin\rfloor-1)}{\Nt},
\end{align}
we rewrite the entries of $\bT$ and $\bC$ as
\begin{align}
    t_\ell & =  e^{j\theta_\ell} \frac{y_\ell}{x_\ell} \\
    c_\ell & = e^{j(\theta_\ell+\Delta \theta_\ell)} \frac{y_\ell+\Delta y_\ell}{x_\ell+\Delta x_\ell}.
\end{align}
Before examining this squared difference in (\ref{FMA}), we introduce the proceeding lemma.

\begin{lemma}
\label{lemma2}
The quantities defined above satisfy
\begin{align}
    \frac{2\eta|\ell|}{\Nmax}\leq&|x_\ell|,|x_\ell+\Delta x_\ell|\leq\frac{\pi\eta|\ell|}{\Nmax} \label{eq:x}\\
    |\Delta x_\ell|\leq \frac{1}{6} & \left(\frac{\pi\eta|\ell|}{\Nmax}\right)^{\!3} \label{MWC} \\
    |y_\ell|,|y_\ell+\Delta y_\ell| & \leq 1\label{eq:y}\\
   |\Delta y_\ell| & \leq\frac{\pi|\ell|}{\Nt} \label{salut} \\
    |\Delta \theta_\ell| & \leq \frac{2\pi|\ell|}{\Nt}
\end{align}
for $|\ell|\leq \Nt/2$, and
\begin{align}
    \frac{1}{x_\ell^2},\frac{1}{(x_\ell+ \Delta x_\ell)^2} \leq \left( \frac{\Nmax}{2\eta\Nt} \right)^{\!\!2} \! \left( \frac{1}{(|\ell|/\Nt)^2}\!+\!\frac{1}{(1-|\ell|/\Nt)^2} \right)\label{eq:fracX}
\end{align}
for $|\ell| < \Nt$. We note that (\ref{eq:x}), (\ref{eq:y}), and (\ref{eq:fracX}), compactly indicate inequalities that apply  to two distinct quantities.
\end{lemma}
\begin{IEEEproof}
Eq. \eqref{eq:x} can be directly derived from
\be
\frac{2}{\pi}|x|\leq |\sin x| \leq |x| \qquad\quad -\pi/2\leq x\leq \pi/2,
\ee
while (\ref{MWC}) follows from
\begin{align}
    |\Delta x_\ell| &= \left|\sin\! \frac{\pi\eta\ell}{\Nmax}-\frac{\eta \Nt}{\Nmax}\sin\! \frac{\pi\ell}{\Nt}\right|\\
    &\leq
    \left|\sin\! \frac{\pi\eta\ell}{\Nmax}-\frac{\pi\eta\ell}{\Nmax}\right|\\
    &\leq \frac{1}{6}\left(\frac{\pi\eta|\ell|}{\Nmax}\right)^3
\end{align}
where the last inequality descends from Taylor's theorem. In turn, \eqref{eq:y} follows from
\be
|\sin y|\leq 1
\ee
and (\ref{salut}) from
\begin{align}
    |\Delta y| &=\left|\sin\! \frac{\pi\eta \ell N_{\rm r}}{\Nmax}-\sin\! \frac{\pi\ell\lfloor\eta\Nmin\rfloor}{\Nt}\right|\\
    &\leq \left| \frac{\pi\eta \ell N_{\rm r}}{\Nmax}- \frac{\pi\ell\lfloor\eta\Nmin\rfloor}{\Nt}\right| \label{eq:mvt}\\
    &\leq\frac{\pi|\ell|}{\Nt}
\end{align}
where, in \eqref{eq:mvt}, the mean value theorem has been applied. Then,
\begin{align}
    |\Delta \theta_\ell|&=\pi|\ell|\left|\left(\!\frac{\eta\Nr}{\Nmax}-\frac{\floor{\eta\Nmin}}{\Nt}\!\right)+\left(\!\frac{1}{\Nt}-\frac{1}{\Nmax}\!\right)\right|\\
    &\leq \pi|\ell|\left(\frac{2}{\Nt}-\frac{1}{\Nmax}\right) \label{eq:theta}\\
    &\leq \frac{2\pi|\ell|}{\Nt}
\end{align}
where \eqref{eq:theta} derives from the triangle inequality.
Finally, \eqref{eq:fracX} can be obtained from
\begin{align}
    \frac{1}{\sin^2 x} \leq \frac{\pi^2}{4}\left(\frac{1}{|x|^2}+\frac{1}{(\pi-|x|)^2}\right) \qquad\quad -\pi\leq x\leq \pi .
\end{align}
\end{IEEEproof}

Utilizing the Toeplitz structure of $\bT$ and $\bC$, we compute the squared difference in \eqref{FMA} for three cases: 1) $\ell=0$, 2) $|\ell|\leq\Nt/2$, and 3) $|\ell|>\Nt/2$. In case of $\ell=0$ where both $t_\ell$ and $c_\ell$ are indeterminate form, $|t_\ell-c_\ell|^2$ is bounded by a constant:
\begin{align}
    \left(\Nr-\frac{\floor{\eta\Nmin}}{\eta\Nt/\Nmax}\right)^2\leq \left(\frac{\Nmax}{\eta\Nt}\right)^2.
\end{align}

For $0<|\ell|\leq \Nt/2$, 
\begin{align}
    &|t_\ell-c_\ell|^2=\!\left|
    \frac{y_\ell}{x_\ell}-e^{j\Delta\theta_\ell}\frac{y_\ell+\Delta y_\ell}{x_\ell+\Delta x_\ell}
    \right|^2\\
    & \quad =\!\left(\!
    \frac{y_\ell}{x_\ell}-\frac{y_\ell+\Delta y_\ell}{x_\ell+\Delta x_\ell}
    \!\right)^{\!\!2}
    \!\!+\!4\sin^2 \frac{\Delta\theta_\ell}{2}\cdot\frac{y_\ell}{x_\ell}\frac{y_\ell+\Delta y_\ell}{x_\ell+\Delta x_\ell}.
\end{align}
Using Lemma \ref{lemma2} and the triangle inequality, the first term can be further bounded by a constant:
\begin{align}
    \left|\frac{y_\ell}{x_\ell}-\frac{y_\ell+\Delta y_\ell}{x_\ell + \Delta x_\ell}\right|
    &\leq \left|\frac{y_\ell\Delta x_\ell}{x_\ell(x_\ell+\Delta x_\ell)}-\frac{\Delta y_\ell}{x_\ell+\Delta x_\ell}\right|\\
    &\leq \left|\frac{y_\ell\Delta x_\ell}{x_\ell(x_\ell+\Delta x_\ell)}\right|+\left|\frac{\Delta y_\ell}{x_\ell+\Delta x_\ell}\right|\\
    &\leq \frac{\frac{1}{6} (\pi\eta|\ell|/\Nmax)^3}{(2\eta|\ell|/\Nmax)^2}+\frac{\pi|\ell|/\Nt}{2\eta|\ell|/\Nmax}\\
    &=\frac{\pi\eta\ell}{6\Nmax}+\frac{\pi\Nmax}{2\eta\Nt}\\
    &\leq \frac{\pi\eta\Nt}{12\Nmax}+\frac{\pi\Nmax}{2\eta\Nt}.
\end{align}
The second term also is bounded by a constant:
\begin{align}
    \left|4\sin^2 \frac{\Delta\theta_\ell}{2}\cdot\frac{y_\ell}{x_\ell}\frac{y_\ell+\Delta y_\ell}{x_\ell+\Delta x_\ell}\right|\!&\leq\!\left| (\Delta\theta_\ell)^2\frac{y_\ell}{x_\ell}\frac{y_\ell+\Delta y_\ell}{x_\ell+\Delta x_\ell}\right|\\
    &\leq \left(\frac{\pi\Nmax}{\eta\Nt}\right)^2.
\end{align}
Altogether, for $|\ell|\leq \Nt/2$ the aggregate squared difference  satisfies
\begin{align}
\sum_{|\ell|\leq \Nt/2}(\Nt-|\ell|) \, |t_\ell-c_\ell|^2=\mathcal{O}(N^2_{\rm t}) .
\end{align}

Proceeding to $|\ell|>\Nt/2$, we begin by noting that
\begin{align}
    |t_\ell-c_\ell|^2\leq 2 \, (|t_\ell|^2+|c_\ell|^2).
\end{align}
Then, using again Lemma \ref{lemma2}, we have that
\begin{align}
    &\sum_{|\ell|>\Nt/2} \!\! (\Nt-|\ell|) \, |t_\ell|^2\\
    &\leq \left(\frac{\Nmax}{2\eta\Nt}\right)^{\!\!2} \left(\sum_{|\ell|>\Nt/2} \!  \frac{\Nt-|\ell|}{(|\ell|/\Nt)^2} + \!\!\! \sum_{|\ell|>\Nt/2} \! \frac{\Nt-|\ell|}{(1-|\ell|/\Nt)^2}\right) \! .
\end{align}
where the first term satisfies
\begin{align}
    \left(\frac{\Nmax}{2\eta\Nt}\right)^{\!\!2} \! \sum_{|\ell|>\Nt/2} \! \frac{\Nt-|\ell|}{(|\ell|/\Nt)^2} = \mathcal{O}(N^2_{\rm t})
\end{align}
while the second term satisfies
\begin{align}
    \left(\frac{\Nmax}{2\eta\Nt}\right)^{\!\!2} \!\! \sum_{|\ell|>\Nt/2} \! \frac{\Nt-|\ell|}{(1-|\ell|/\Nt)^2} & =\left(\frac{\Nmax}{2\eta}\right)^{\!\!2} \!\! \sum_{|\ell|>\Nt/2} \! \frac{1}{\Nt-|\ell|} \nonumber \\
    &=\mathcal{O}(N^2_{\rm t} \log \Nt),
\end{align}
with the last equality holding by virtue of the logarithmic growth of the harmonic series. The same result can be obtained for $c_\ell$, hence
\begin{align}
    \|\bT-\bC\|_{\rm F}^2&= \! \sum_{|\ell|<\Nt}(\Nt-|\ell|) \, |t_\ell-c_\ell|^2\\
    &= \mathcal{O}(N^2_{\rm t} \log \Nt).
\end{align}

\bibliographystyle{IEEEtran}
\bibliography{ref}

\end{document}